# Influence of structural defects on the magnetocaloric effect in the vicinity of the first order magnetic transition in $Fe_{50.4}Rh_{49.6}$

V. I. Zverev, A.M. Saletsky

Faculty of Physics, M.V. Lomonosov Moscow State University, 119991, Moscow, Russia

R.R. Gimaev[1], A.M. Tishin

Faculty of Physics, M.V. Lomonosov Moscow State University, 119991, Moscow, Russia
Advanced Magnetic Technologies and Consulting LLC, 142190, Troitsk, Russia

T. Miyanaga

Department of Mathematics and Physics, Hirosaki University, Hirosaki, Aomori, 036-8561, Japan

J. B. Staunton
Department of Physics, University of Warwick, Coventry CV4 7AL, U.K.

---

[1] Corresponding author. Electronic mail gimaev@physics.msu.ru

**ABSTRACT**

The large magnetocaloric effect (MCE), which accompanies the first order ferromagnetic/anti-ferromagnetic transition in CsCl-ordered Fe-Rh alloys, has been investigated by measurements in slowly cycled magnetic fields of up to 2 T in magnitude for a range of temperatures, 300K < T < 350K. A bulk sample with composition $Fe_{50.4}Rh_{49.6}$ was used and the results were compared with those produced by the ab-initio density functional theory-based disordered local moment (DLM) theory of the MCE. The measurements revealed an irreversibility effect in which the temperature of the material did not return to its initial value following several cycles of the magnetic field. These observations were explained in the framework of the ab-initio theory for the first order transition in which the consequences of the incomplete long range compositional order and small compositional inhomogeneities of the sample were included. The mean value of the long range order parameter S used in the theoretical work was 0.985, close to the value obtained experimentally from XRD measurements. The sample inhomogeneities were modeled by regions in the sample having a distribution of S values with narrow half-width 0.004 about the mean value. The influence of such compositional disorder on both the transition temperature (323.5 K) and MCE adiabatic temperature change ($\Delta T$ = 7.5 K) was also studied.

**INTRODUCTION**

Magnetic refrigeration has an important role in the development of efficient and environmentally friendly solid state cooling technologies. At its heart is the magnetocaloric effect (MCE) which leads to an adiabatic temperature change ($\Delta T$) and isothermal magnetic entropy change ($\Delta S_{magn}$) in magnetic materials when an external magnetic field is applied. As well as being used in refrigerators, materials showing a large MCE have potential application in magnetic heat engines and also medicine [1-3]. There is currently much research activity in the search for materials which show large values of $\Delta T$ and $\Delta S_{magn}$ in modest magnetic fields with the expectation that it should be possible to find one with a $\Delta T$ as large as ~18 K/T [4].

In approximately equal proportions iron and rhodium produce an ordered alloy structure which undergoes a first order transition between a ferromagnetic and antiferromagnetic phase around room temperature. This transition is accompanied by a large MCE. The alloy with $Fe_{0.49}Rh_{0.51}$ composition possesses the record for measured MCE values of all magnetocaloric materials studied to date [2,5-7]. This giant MCE is accompanied by magneto-structural changes. From this perspective, following some recent studies of the magnetocaloric properties of FeRh[8-

[11], it is interesting to conduct a more thorough experimental and theoretical study of the alloy to try to understand the cause of its large MCE and to gain useful insight for the search for other MCE materials

In 1938, M. Fallot[12,13] first showed that, with increasing temperature, the ordered magnetic alloy $Fe_{50}Rh_{50}$ undergoes an isostructural [14] phase transition from an antiferromagnetic type I (AFM) phase to a ferromagnetic phase (FM) at a temperature $T_{tr} \sim 320$ K, which is accompanied by an increase in volume of approximately 1% [15–17]. The measured $T_{tr}$ exhibits hysteresis and there is a temperature region where the AFM and FM phases coexist [18]. The $Fe_{50}Rh_{50}$ ordered alloy has a B2 crystal structure (type CsCl) [13,14]. It is known that collinear ordering of magnetic moments on Fe (3.2 $\mu_B$/atom) and Rh (0.9 $\mu_B$/atom) atomic sites exists in the ferromagnetic phase [19,20]. At lower temperatures the AFM phase appears in which the Fe atoms form two magnetic sublattices with oppositely directed magnetic moments. The magnetic moments associated with the Fe atoms in the AFM phase are 3.3 $\mu B$ [21] whereas there are no magnetic moments on the Rh atoms [22-24].

Further ab-initio theory considerations [25] show that a very slight change in either the stoichiometry or long-range B2 order has a strong influence on the FM/AFM phase transition. In particular, such compositional variations, as observed in real materials, lead to small percentages of the Rh atoms being substituted by Fe atoms on the Rh sub-lattice. The transition in FeRh is manifested as a result of a delicate balance of competing electronic influences which is disrupted by the slightest changes in composition. The conclusion is that this hypersensitivity to such compositional variations causes problems for the applications of this alloy as a technologically useful magnetic material.

Despite the large number of works devoted to the origin of the first order magnetic phase transition in FeRh[8-11], there is still discussion about the characteristic times of the processes associated with the structural and magnetic sub-lattices as well as their interplay in the vicinity of the phase transition. Some authors propose that a change in volume leads to the magnetic transition [26,27], whilst other works conclude that the magnetic transition is simply accompanied by a volume change. In this letter we describe our study of the dynamics of the MCE in FeRh. This is of key relevance for magnetic refrigeration where the magnetic field is repeatedly cycled, in controllable drug release technologies [33] and in magnetic memory applications [34]. Substantial attention to the irreversible processes which are observed during the demagnetization/remagnetization of these alloys is required.

In this paper, we report our experimental verification of theoretical predictions concerning the influence of incomplete B2 long range order (partial replacement of Rh atom positions by Fe and vice versa [25]) on both $T_{tr}$ and the maximum MCE values. We have also found a theoretical explanation of an irreversibility effect which we have observed experimentally where we discovered that the temperature of the material did not return to its initial value following the first cycle of MCE measurements.

## THEORETICAL CONSIDERATION

Our starting point was the ab-initio disordered local moment theory [25,35] for materials with quenched static compositional order. We specified a Fe-Rh alloy for any composition close to the stoichiometric, perfectly ordered B2 phase as Fe(100-$x$)Rh($x$) – Rh(100-$y$)Fe($y$) where both $x$ and $y$ are small percentages. In these terms the Fe proportion of the material $c = (100-x+y)/2$ and the long-range order parameter $S = 1-y/100$. For a specific composition, $c$, $y$, the free energy of such an alloy can be written

$$F(c,y,H,T) = U(c,y,m_f,m_a) - T(S_{mag}(m_f,m_a,T) + S_{latt}(T)) - Hm_f \qquad (1)$$

where $U$ is the internal magnetic energy, $S_{mag}$ is the magnetic entropy, $S_{latt}$ the lattice vibration entropy (from a Debye model), H is magnetic field and $m_f$ is the ferromagnetic order parameter (proportional to the overall magnetization) and $m_a$ is the anti-ferromagnetic order parameter. For complete AFM order $m_a=1$, $m_f=0$, for complete FM order, $m_a=0$, $m_f=1$ and in the high temperature paramagnetic state both are zero. From our ab-initio calculations for several explicit $x$ and $y$ values [18] we found that the internal energy

$$U(c,y,m_f,m_a) = -\left(e_f m_f^2 + e_{af} m_a^2 + g_f m_f^4 + g_{af-f} m_f^2 m_a^2 + g_a m_a^4\right) \qquad (2)$$

where (in meV) $e_f = 100 + 4.80(c-50) + 6.94y$, $e_a = 88 - 1.05(c-50) + 2.8y$, $g_f = -23$, $g_{af-f} = 18$ and $g_a = 36$. The values of $m_f$ and $m_a$ were taken from where the free energy of Eq.(1) was minimized. As shown previously [25] we obtained a good qualitative description of the first order FM-AFM phase transition and its accompanying large MCE in FeRh. We used this model to analyze the experimental results reported in this paper.

## EXPERIMENTAL TECHIQUE

A FeRh alloy was prepared by the plasma arc melting method (PAM). The sample cell was evacuated to $10^{-3}$ Pa and the substituted Ar gas pressure was 0.09 MPa. The FeRh sample was annealed at 1273 K for 48 hours in vacuum and quenched by cooling into water. The chemical composition of the alloy was determined by an electron probe micro analyzer EPMA (JEOL JXA-8800RL, JEOL ltd) and evaluated to be $Fe_{50.4}Rh_{49.6}$. It was primarily in the B2 ordered phase with a very small proportion of the disordered fcc phase which shows no FM/AFM transition and has insignificant MCE. The structure of the prepared bulk slice of $Fe_{50.4}Rh_{49.6}$ alloy was analyzed by XRD using the Cu $K_\alpha$ line from the X-ray diffractometer (M18XHF-SRA, MAC Science co. ltd.). Magnetization measurements were performed using a vibrating sample magnetometer (VSM, Toei co ltd. model-5) with a sweep rate of 1 K/min.

The magnetic field dependence of the MCE was measured at different temperatures in the region of the first order AFM–FM phase transition of the $Fe_{50.4}Rh_{49.6}$ bulk sample, and the temperature dependence of the MCE in the largest magnetic field of 1.8 T was obtained by direct measurement using the automated MCE measuring setup (MagEq MMS 801, AMT&C LLC). A detailed description of the equipment and measuring method has been presented in [36]. All the temperature measurements were performed by direct measurements. We determined the FM/AFM transition temperature from where the MCE reached its maximum value [37,38].

To avoid the influence of the thermal hysteresis and to provide repeatability of the results the sample was cooled well into the AFM state to 270 K before each measurement of the MCE. A similar scheme of direct measurements has been applied previously [6].

**RESULTS AND DISCUSSION**

Figure 1 shows the measured XRD pattern for $Fe_{50.4}Rh_{49.6}$. For such a measurement system, the measured lattice constant $a$ is expressed in terms of the actual lattice constant $a_0$, empirical coefficient $K$ and scattering angle $\theta$ : $a = a_0 + a_0 K \cos^2\theta$. By extrapolation, $a_0$ was determined as 2.991 ± 0.004 Å.

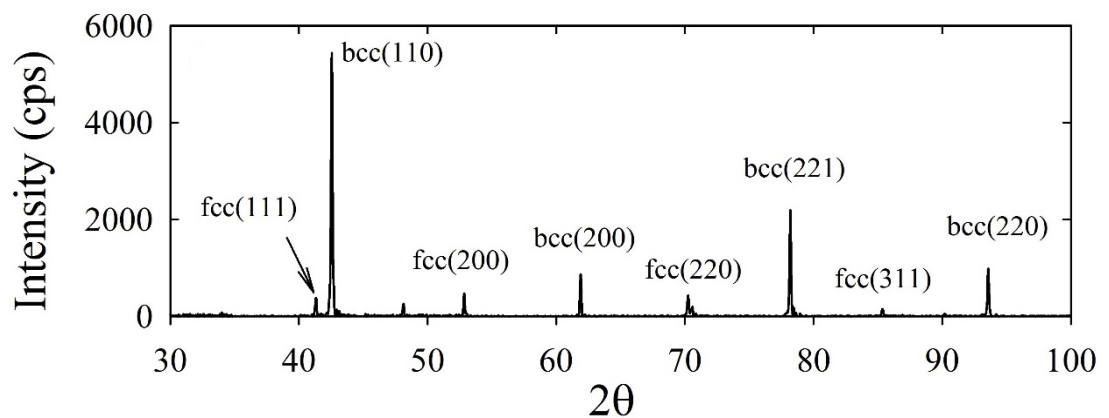

Fig. 1. XRD results for the $Fe_{50.4}Rh_{49.6}$ sample.

From the measured XRD intensity the long range order parameter $S$ was estimated using following equation:

$$S = \frac{f_{Fe}+f_{Rh}}{f_{Fe}-f_{Rh}}\sqrt{\frac{I_s}{I_f}} \qquad (3)$$

where $f_{Fe}$ and $f_{Rh}$ were the atomic scattering factors for the Fe and Rh atoms respectively, and $I_S$ and $I_f$ were the scattering intensities for the superlattice and fundamental lines respectively. The value of $S$ was found to be 0.985, i.e. where just 1.5% of the positions of the Fe atoms on the Fe sub-lattice were substituted by Rh atoms and vice-versa.

Figure 2 shows the temperature dependence of the magnetization, the $M$–$T$ curve, for $Fe_{50.4}Rh_{49.6}$ over the range of 0.25~1.0 T. The inset shows the measured $dM/dT$ versus $T$ curve at 1 T. The temperature hysteresis was 12 K in this case. For higher magnetic fields a lower $T_{tr}$ was observed.

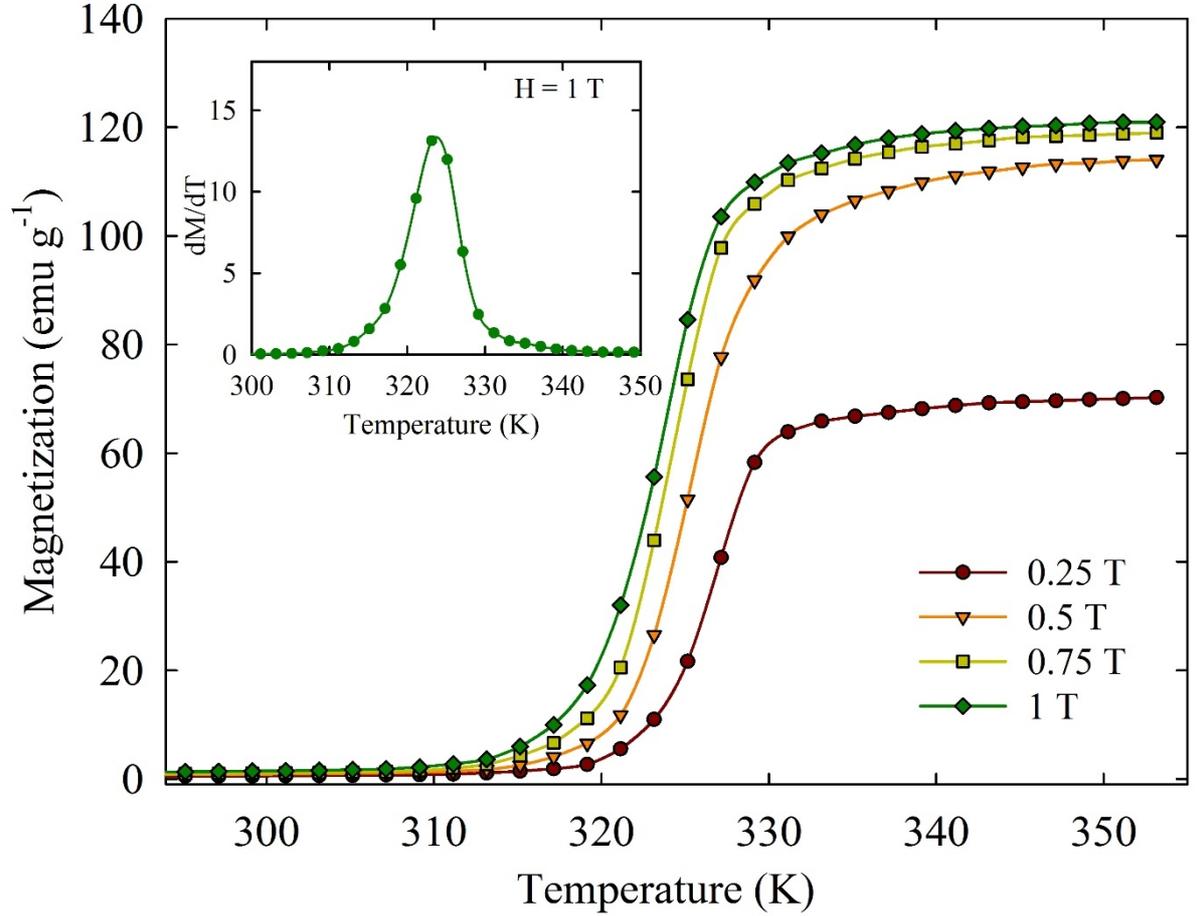

Fig. 2. $M$–$T$ curves for $Fe_{50.4}Rh_{49.6}$, which were obtained while heating the sample from the AFM state in magnetic fields of 0.25~1.0 T. The inset shows the temperature dependence of dM/dT at 1 T.

Fig. 3(a) shows the temperature dependence of the $\Delta T$ in a magnetic field 1.8 T. The measurements showed the alloy to exhibit a negative MCE with a maximum at about 323.5 K which was in a good agreement with the magnetic measurements (see inset of Fig. 2).

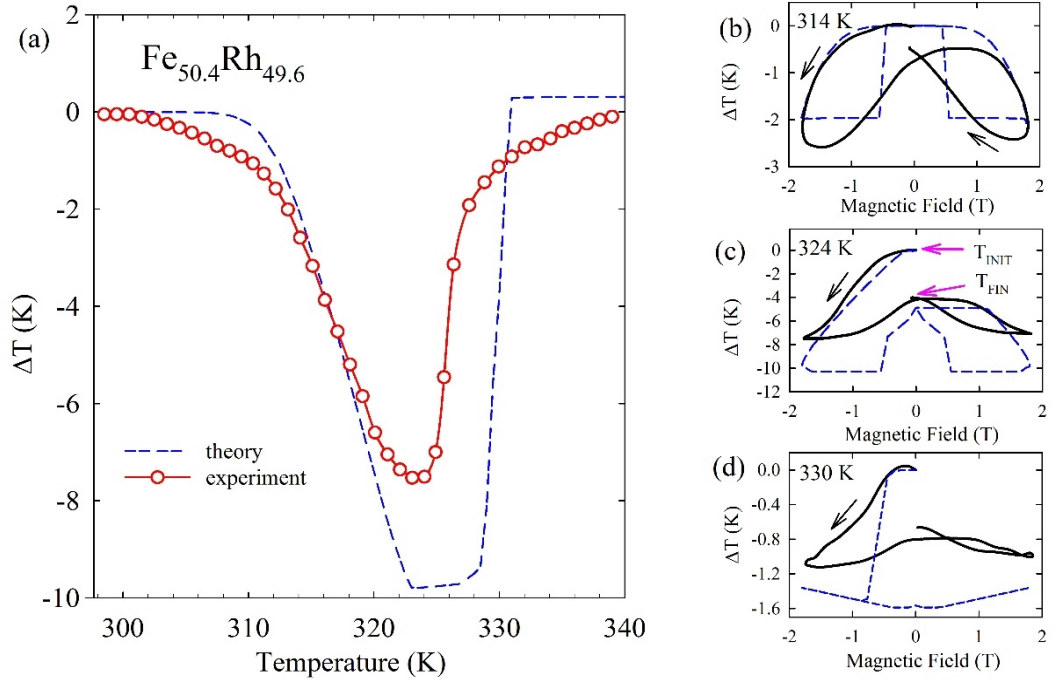

Fig. 3. (a) ΔT(T) for the Fe$_{50.4}$Rh$_{49.6}$ in a magnetic field of 1.8 T and ΔT(H) for one full cycle of the magnetic field at temperatures (b) 314 K, (c) 324 K and (d) 330 K. Black arrows indicate the direction in which the temperature was changed, the blue arrows in (c) indicate the temperature of the sample at the initial ($T_{INIT}$) and end ($T_{FIN}$) times for the application of the field. The dashed lines show the results from the theoretical model.

Typical ΔT(H) curves are shown in Figure 3(b) – (d): (b) below $T_{tr}$ (314 K), where a significant MCE value was observed; (c) in the vicinity of $T_{tr}$ (324 K), where the MCE reached its maximum value, and (d) above $T_{tr}$, where a MCE was still observed (330 K). Before measurement the sample was in a zero magnetic field at $T_{INIT}$. Then once the magnetic system had been launched, the magnetic field was increased in size at a rate of 1 T/s to -1.8 T, which caused a decrease in temperature owing to the negative MCE. The magnetic field was subsequently reduced (in magnitude) to zero, which was accompanied by an increase in the temperature of the sample. After that the field was increased again towards positive values causing the temperature of the sample to drop. Upon reaching the maximum value of 1.8 T, the magnetic field was reduced to zero again (at temperature $T_{FIN}$).

We used the ab-initio disordered local moment model to understand these experimental findings. We set the average Fe concentration, *c*, to be that of the sample, 50.4%. We assumed that the sample had some tiny fluctuations in its composition so that it comprised many regions characterized by slightly different values of long range order, *S* = 1 - *y*, close to one. Values were taken from a distribution with Gaussian weighting, i.e. with a probability P(S)= (1/w)exp[-(S-S$_m$)$^2$/w$^2$], where the half-width w=0.004 and the mean value of the long range order S$_m$= 0.985.

We calculated $\Delta T$ for each value of $S$ and then averaged over compositions using the Gaussian weighting. The choice of the mean $S_m$ as 0.985 (similar to the experimental estimate of 0.985) brought the calculated $\Delta T$ peak in 1.8 T close to the experimental values as shown in Fig 3(a), although the theoretical magnitude was 20 % bigger. This overestimate was likely caused by the mean field basis of the theory's neglect of fluctuations. The sharper reduction of $\Delta T(T)$ outside the transition region 310 K to 330 K found in the theory also had the same cause.

We then sought to find a simple mechanism that could reproduce the observed results when the applied magnetic field was cycled. We assumed that there was a weak, slow, long-ranged dynamic magnetizing effect coming from the small compositional inhomogeneities of the sample. Each spatial region with its specific composition S was affected by the overall magnetization from the rest of the sample which we modeled in a very simple way. The time variation of the field was sufficiently slow so that equilibrium thermodynamics could be applied at every time step, $t_i$, and the extent of ferromagnetic or anti-ferromagnetic order, $m_f(t_i)$ and $m_a(t_i)$, averaged over compositional variations and $\Delta T$ for each value of $H(t_i)$ could be found. This dynamic effect led to an extra effective magnetic field of form, $H_{eff} = Cm_f(t_{i-1})$, being added to the external field $H(t_i)$. $H_{eff}$ depended on the overall magnetization integrated over earlier times and $C$ was a phenomenological parameter. This simple dynamic effect added into the theoretical model produced qualitatively the effects found in the experiments. Theoretical curves in Figures 3(b), 3(c) and 3(d) show the theoretical $\Delta T(H)$ as $H(t)$ was cycled and are directly comparable to the experimental curves.

The experimentally measured $\Delta T$ as a function of applied field H revealed hysteresis as the field was varied (FWHM of about 1.2 T) and had a characteristic feature showing an "irreversibility" of the sample's temperature to the initial temperature after one complete cycle of the magnetic field (as marked with purple arrows in Fig. 3(d)). The "irreversibility" effect was observed for all temperatures in the range 300 – 340 K, and the value ($T_{FIN} - T_{INIT}$) depended on temperature, which is shown in Fig. 4(a). For comparison, the same figure shows the temperature dependence of $\Delta T(T)$, presented earlier in Figure 3(a). The "irreversibility" (finite ($T_{FIN} - T_{INIT}$) value) reached a peak value at 325 K. It is noticeable that the temperature where ($T_{FIN} - T_{INIT}$) was maximum was shifted up by ~ 2 K in comparison with the $\Delta T(T)$ peak.

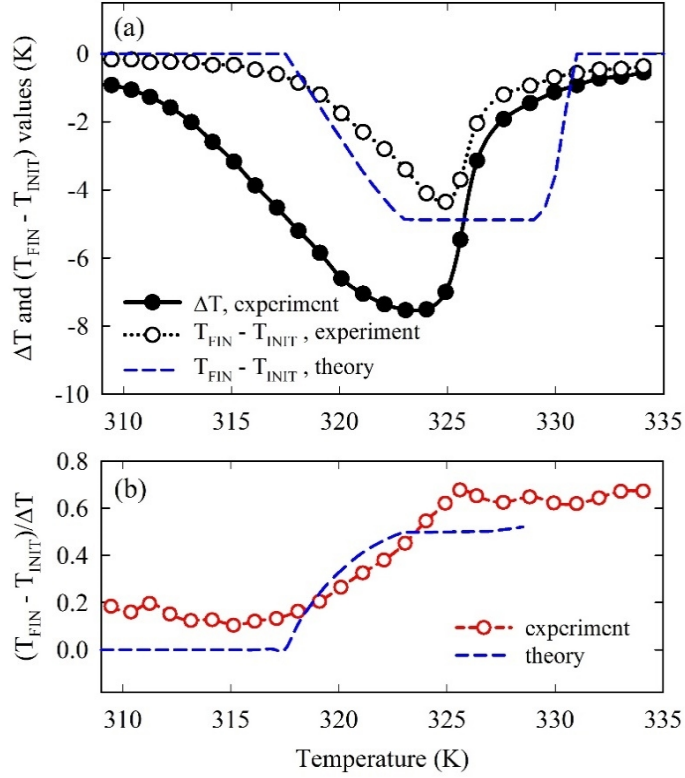

Fig. 4. Temperature dependence of (a) the absolute (open circles) and (b) the relative values of ($T_{FIN}$ - $T_{INIT}$) for the $Fe_{50.4}Rh_{49.6}$ from both experiment and theory. The plot (a) also shows the experimental $\Delta T(T)$ dependence for a comparison of the peaks' positions.

The temperature dependence of the relative value ($T_{FIN}$ - $T_{INIT}$)/$\Delta T$ is shown in Fig. 4(b). It is evident that in the range 300–340 K, the relative value of the 'irreversibility' did not change up to 315 K and a further increase of temperature led to a monotonic increase of relative "irreversibility" to 0.65 at 326 K. At temperatures above 326 K ($T_{FIN}$ - $T_{INIT}$)/$\Delta T$ did not change. Fig. 4(a) also shows the theoretical ($T_{FIN}$ - $T_{INIT}$) temperature dependence. The phenomenological parameter C, described above, was chosen to be 3.8 T to bring the theoretical values of ($T_{INIT}$ − $T_{FIN}$) to a similar size to the experimentally measured values. We found that the theory model gave a qualitative explanation of the experimental measurements. It tracked the temperature dependence of $\Delta T$ as shown in Fig. 4(b) (dashed line) and its peak value was roughly a half of the maximum in $\Delta T$. One can conclude that the first order magnetic phase transition observed in FeRh was the reason for the irreversibility effect since it is explicitly included in the theory model and such behavior is not observed in magnetic materials like Gd which undergo second order phase transitions [36]. We also note that both $\Delta T$ and ($T_{FIN}$ − $T_{INIT}$) had a similar non-symmetric temperature dependence as observed in the experimental results. This can be understood from the following considerations. $\Delta T(T)$ had the largest magnitude just below $T_{tr}$ where the free energy (Eq.1) of the FM phase was equal to that of the AFM phase, i.e. $\Delta F = F_{FM}(H = 0, T_{tr}) - F_{AFM}(H = 0, T_{tr}) = 0$. Application of a magnetic field, $H$, changed the AFM order to FM order and a large magnetic entropy change occurred with large $\Delta T$. As $T$ was lowered $\Delta T(T)$ decreased to zero - $\Delta F$ increased as the AFM phase became more entrenched. Eventually both ($F_{FM}(H \neq 0,T) - F_{AFM}(H = 0,T)$) and $\Delta F > 0$ so that application of the field no longer promoted a change from antiferromagnetic order and $\Delta T \approx 0$.

Measurements at higher speeds (up to 6 T/s) of the magnetic field sweep showed that both $\Delta T$ and ($T_{FIN} - T_{INIT}$) did not depend on the rate of field change. Fig. 5 shows the $\Delta T(H)$ plot for three full cycles of the magnetic field at 323 K, close to $T_{tr}$. During the first cycle $\Delta T_1$ reached -7.5 K and during the 2$^{nd}$ and 3$^{rd}$ cycles the MCE values were about two times smaller (-3.4 K). The figure prompts the suggestion that the decrease in $\Delta T$ in the 2$^{nd}$ and 3$^{rd}$ cycles comes from the "irreversibility" effect. This is supported by the theoretical modeling (dashed line in Fig. 4(b)) which found a similar behavior. During the first cycle $\Delta T_1$ reached ~ -9.8 K and in second and subsequent cycles the $\Delta T_2$, $\Delta T_3$,... MCE values were ~ -4.9 K, roughly two times smaller.

From the practical point of view this observation implies lower cooling efficiency in multiple magnetization/demagnetization cycles during the operation of a refrigerator. The presence of field hysteresis of the MCE curves should also be taken into account. Thus, the elucidation of the origin of the one-time "irreversibility" of the sample temperature during the first cycle and the almost two-fold decrease of the MCE during subsequent cycles has importance for the development of a material with large magnetocaloric properties, which persist during multiple cycles of magnetic field.

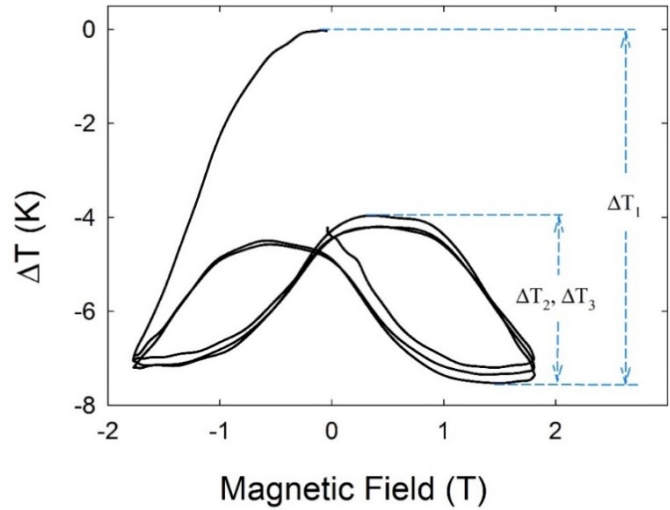

Fig. 5. $\Delta T(H)$ obtained over three cycles of magnetization/demagnetization at 324 K for the Fe$_{50.4}$Rh$_{49.6}$ sample.

**CONCLUSION**

We have presented results of experimental investigations of the crystallographic structure, magnetization and MCE of a $Fe_{50.4}Rh_{49.6}$ bulk sample. The experimental results have been theoretically described in the framework of the ab-initio disordered local moment theory model. It has been shown that slight variation from complete compositional B2 order and small compositional inhomogeneities significantly influence both the AFM-FM phase transition temperature and the behavior of the temperature dependence of magnetocaloric properties. The comparison of theoretically determined and experimentally measured MCE values supported the theoretical conclusion that there is a large electronic contribution (up to 40 %) during the AFM - FM transition in FeRh alloys. We have also observed and explained the reason for a sharp decrease and hysteresis of MCE (almost two-fold) during demagnetization/remagnetization cycles in the material.

**ACKNOWLEDGEMENTS**

Work in Advanced Magnetic Technologies and Consulting, LLC was supported by Skolkovo Foundation, Russia. T.M. thanks Prof. T.Okazaki and Mr. M.Ohno for the preparation and characterization of the sample. J.B.S acknowledges support from EPSRC (UK) Grant No. EP/J006750/1.